\documentclass[a4paper,fleqn,usenatbib]{mnras}
\usepackage[T1]{fontenc}
\usepackage{ae,aecompl}
\usepackage{multirow,ulem}

\usepackage{graphicx}	
\usepackage{amsmath}	
\usepackage{amssymb}	
\usepackage{natbib}
\usepackage{float}
\usepackage{appendix}
\usepackage[dvipsnames]{color}
\definecolor{webgreen}{rgb}{0,.5,0}
\title[CGM in the photoionized precipitation model]{{A panoramic view of the circumgalactic medium in the photoionized precipitation model}}
\author[Roy, Nath, Voit]
{
Manami Roy$^{1}$
\thanks{E-mail : manamiroy@rri.res.in },
Biman B. Nath$^{1}$, G. M. Voit$^{2}$ \\
\footnotesize \it $^{1}$Raman Research Institute, Sadashiva Nagar, Bangalore 560080, India \\ 
\footnotesize \it $^{2}$Department of Physics \& Astronomy, Michigan State University, East Lansing, MI 48824, USA \\
}

\date{Accepted XXX. Received YYY; in original form ZZZ}

\pubyear{2020}

\begin{document}
\label{firstpage}
\pagerange{\pageref{firstpage}--\pageref{lastpage}}
\maketitle


\begin{abstract}
{We consider a model of the circumgalactic medium (CGM) in which feedback maintains a constant ratio of cooling time to freefall time throughout the halo, so that the entire CGM is marginally unstable to multiphase condensation. This `precipitation model' is motivated by observations of multiphase gas in the cores of galaxy clusters and the halos of massive ellipticals.  We derive from the model density and temperature profiles {for the CGM around galaxies} with masses similar to the Milky Way.  After taking into consideration the geometrical position of our solar system in the Milky Way, we show that the CGM model is consistent with observed {OVI}, OVII and OVIII column densities only if temperature fluctuations with a log-normal dispersion $\sigma_{\ln T} \sim 0.6$--$1.0$ are included.  
We show that OVI column densities observed around star-forming galaxies require systematically greater values of $\sigma_{\ln T}$ than around passive galaxies, implying a connection between star formation in the disk and the state of the CGM.  Photoionization by an extra-galactic UV background does not significantly change these CGM features for galaxies like the Milky Way but has much greater and {significant} effects on the CGM of lower-mass galaxies.} \\

\end{abstract}




\section{Introduction}
{The circumgalactic} medium (CGM) {is} a reservoir {of diffuse gas} that fills the region surrounding the optically visible part of a galaxy, extending up to its virial radius 
\citep{Tumlinson2017}. It has been inferred that the CGM contains a substantial amount of baryons, thereby possibly mitigating the problem of missing baryons \citep{Tumlinson2017}. This gaseous halo {may date} back to the epoch of formation of the host galaxy, but its properties have likely evolved  due to various feedback processes \citep{White1991}. The CGM is believed to play a big role in the evolution of a galaxy by supplying gaseous fuel that triggers star formation  \citep{Keres2005, Keres2009,Sancisi2008}, by acting as the `waste dump' for  outflows of gas, and also by helping to recycle this gas \citep{Tumlinson2017}.

The density, temperature, metallicity structure, ionization state, and total mass of the CGM are yet to be determined with accuracy. These open questions are being addressed through both analytical and semi-analytical theoretical models, as well as numerical simulations.  However, the relevant length scale for studying some of the processes {such as cooling, metal mixing and transport, feedback} that determine the physical state of the CGM is a few tens of parsecs and is difficult to achieve in a numerical simulation also containing the virial radius of a Milky Way-sized galaxy. {Alternatively, theoretical modelling offers a simple, flexible, physically motivated and efficient approach to exploring those issues.} In this regard, \cite{Faerman2017} considered an isothermal model with constant metallicity and two phases of gas (warm and hot) with a log-normal distribution of dispersion $\sigma_{\ln T}$ around each of those (constant) mean temperatures. They concluded that a metallicity of $0.5\ \text{Z}_{\odot}$ and $\sigma_{\ln T}$ of $0.3$ are needed  to match the OVI column density observations. \cite{Qu2018} have recently built an analytical model with $0.3$ Z$_{\odot}$, {based on} balancing the star formation rate and cooling rate, {that predicts CGM} column densities for OVI, OVII, and OVIII. They compared the calculated column densities at 0.3 R$_{\rm vir}$ with observed values and could not find a suitable match. Then they modified the model {to have a} metallicity 0.55 Z$_{\odot}$ and a temperature that is nearly twice the virial temperature in order to match their model values with observations. 
\cite{Faerman2019} have proposed a newer CGM model {that is isentropic} without any temperature fluctuations and in which photoionization {affects the ion abundances 
in such a way that the observed column densities of OVI, OVII and OVIII are reproduced.}  

The existence of lines of different ionization species in the absorption spectra of the CGM strongly points towards a multiphase (temperature and density) structure \citep{Gupta2012,Fang2015}. Various instabilities and radiative cooling cause cool gas to condense out of the hot CGM, leading to a multiphase structure in which there are large and non-radial variations in specific entropy.  Approximate  pressure equilibrium then ensures a multiphase structure in temperature and density. 

{\cite{Field1965} showed} {that local thermal instability is {possible in a diffuse medium in which the} cooling curve has a negative slope and a region with higher density and lower temperature than the surrounding medium {has} formed.} 
Local thermal instability occurs in the CGM when a region of slightly overdense gas starts to cool faster than its surroundings, because the cooling time 
\begin{equation}
    t_{\rm cool} \equiv \frac {3} {2}  
                    \frac {nkT} {n_{\rm H}^2 \Lambda_{\rm N} (T)}
    \; \; ,
\end{equation}
{where $n_{\rm H}^2 \Lambda_{\rm N}$ is the radiative cooling per unit volume, generally decreases with increasing density.} Local pressure equilibrium then implies an isobaric cooling process, causing the density in the clump to increase and the cooling time to decrease further, until efficient radiation emission by atomic lines is not possible ($T \lesssim 10^4$). {However, gravity can interfere with the growth of thermally unstable perturbations, if buoyancy causes the overdense gas to sink faster than it can shed its thermal energy \citep[e.g.,][]{Binney2009}.}

Recent phenomenological studies and numerical simulations {\citep[e.g.,][]{McCourt_2012MNRAS.419.3319M,Sharma2012,Voit2015}} show that there is a threshold value for the ratio of radiative cooling time to free fall time, below which {thermally unstable perturbations can proceed into multiphase condensation.} 
Buoyancy strongly suppresses multiphase condensation in stratified media with a cooling time more than an order of magnitude greater than the local free-fall time, $t_{\rm ff} \equiv (2r/g)^{1/2}$.  But numerical simulations show that thermal instability can produce multiphase condensation (also known as precipitation) in stratified media with $t_{\rm cool}/t_{\rm ff} \lesssim 10$, {as long as moderate disturbances can suppress the damping effects of buoyancy \citep[e.g.,][]{Voit2017ApJ...845...80V,Voit2018ApJ...868..102V,Voit2021ApJ...908L..16V}.} Our physically motivated analytical model of the CGM is therefore a `precipitation model' based on assuming  $t_\text{cool} /t_\text{ff} \approx 10$ \citep[e.g.,][]{Voit2019}. 

In this paper, we extend the work of \citep{Voit2019} {by presenting the angular dependance of OVI, OVII and OVIII column densities on Galactic coordinates for a precipitation-limited CGM model and considering the effects of photoionization on a precipitation-limited CGM model.} We study the  {CGM} density and temperature profiles that result from the `precipitation model' in light of the observed column densities of highly ionized oxygen lines through the CGM of the Milky Way (MW), as well as those of other galaxies. 
We calculate the column density for a general line of sight and show the variation with Galactic latitude and longitude. We consider both collisional ionization and photoionization for calculating the ionization fractions of OVI, OVII, and OVIII.  We also extend our analysis to low mass galaxies {to study the effect of photoionization on the precipitation-limited CGM model around these galaxies.} 
In addition, we incorporate small scale temperature fluctuations at each radius and compare our model with the UV and X-ray observations of these absorption lines by {\it COS-Halos}, {{\it Far-Ultraviolet Spectroscopic Explorer (FUSE)}}, {\it XMM-Newton}, {and the} {\it Chandra} X-ray observatory. 
Lastly, we use our results to compare the column density observations of star forming and passive galaxies, and ask if star formation in the {central galaxy} affects the physical state of {its} CGM.


The structure of the paper is as follows:  Section \ref{Model} describes a  precipitation-limited model for CGM density and temperature {that accounts for photoionization and allows for small-scale temperature fluctuations with a log-normal distribution.}  Section \ref{Results} {applies the model to calculate} column densities of highly ionized oxygen {through the Milky Way's CGM over the full range of} Galactic latitude and longitude. {Section \ref{OBS} compares those results with {UV and} X-ray absorption-line observations of the Milky Way's CGM. Section \ref{Discussion} discusses what those results imply about CGM temperature fluctuations and extends the analysis to OVI observations of other galaxies like the Milky Way, showing star formation is correlated with greater CGM fluctuations.  It also considers the effects of photoionization on precipitation-limited models of the CGM around lower-mass galaxies, showing that including photoionization dramatically increases their OVI column density {as well as affects the column densities of other ions} .  Section \ref{CONCLUSION} summarizes our results.} 
\begin{figure}
  \centering
    \includegraphics[width=0.45\textwidth]{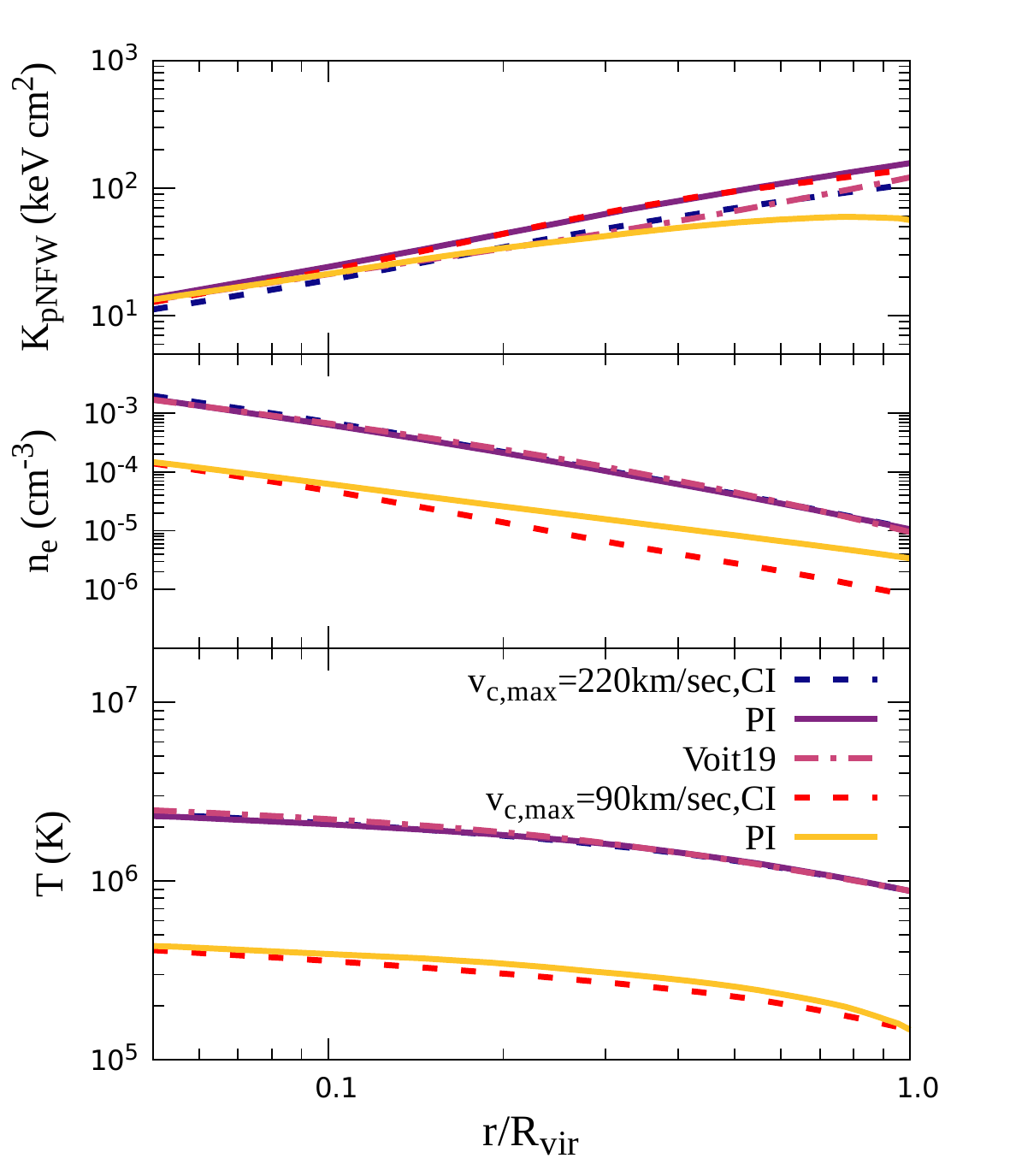}
    \caption{
        {(a) Entropy profile $K_\text{pNFW}$, 
                  (b) the corresponding pNFW electron density profile, 
                  and (c) the pNFW temperature profile}
{for both CI and PI  precipitation-limited model in the case of two          different halo masses (v$_{\rm{c,max}}=90, 220$ km s$^{-1}$).  The profiles of \citet{Voit2019} are also shown for comparison.} 
    \label{profile}}
  \end{figure}

\section{MODELING THE CGM}
\label{Model}
We first present a precipitation-limited CGM model that assumes $t_\text{cool} /t_\text{ff} = 10$ in a galactic potential well.
We assume that the CGM is in hydrostatic equilibrium, so that 
\begin{equation} \label{HSE}
    \frac{dP}{dr}  = -\frac{d\phi}{dr}\rho 
        \; \; \Rightarrow \; \; 
        \frac{dT}{dx} = - \left( \frac{\mu m_p} {k} \right)\frac{d\phi}{dx}  
            - \left( \frac{T}{n} \right)  \frac{dn}{dx} \; , 
\end{equation}
where $\phi$ represents the gravitational potential, $x \equiv r/r_s$ represents radius in units of a scale radius $r_s$, and the other variables have standard meanings.
Following \citet{Voit2019}, we consider a CGM confined by a modified Navarro-Frenk-White (NFW) potential well \citep{NFW}. The circular velocity $v_\text{c} \equiv [ r ( d \phi / dr ) ]^{1/2}$ is constant in the inner part of the potential, with
\begin{equation*}
v_\text{c}(r) = v_\text{c,max} \; \; , \; \; r \leq 2.163 \, r_s \; ,
 \end{equation*}
to account for stellar mass near the center, and {$v_\text{c}$} declines slowly with radius at larger radii, following 
\begin{equation} \label{vc}
    v_\text{c}^2(r) =   v_\text{c,max}^2 (r) 
      \times 4.625 \left[ \frac{\ln(1+r/r_s)}{r/r_s} - \frac{1}{1+r/r_s} \right] \,
\end{equation}
as in a normal NFW profile. This prescription for circular velocity yields the derivative of the potential required in the first term of equation (\ref{HSE}).

At this point, there are several ways to proceed. The simplest is to consider an isothermal model that would render the LHS of equation (\ref{HSE}) zero \citep{Faerman2017, Qu2018}. Another option is to connect temperature and density by specifying the specific entropy\footnote{This paper expresses specific entropy in terms of the entropy index $K \equiv kT n_e^{-2/3}$ usually adopted in CGM studies.} at each radius. One choice is to assume a uniform entropy throughout \citep{Faerman2019}. Our model assumes a composite entropy profile built by combining two physically motivated entropy profiles. The first is the baseline entropy profile produced by non-radiative structure formation, \begin{equation} \label{Kbase}
    K_\text{base} = \left( 39 \, {\rm keV} \, {\rm cm}^2 \right) \,  
        v_{200}^2 \, \left( \frac {r} {r_{200}} \right)^{1.1} \; ,
\end{equation}
taken from \cite{Voit2019}, in which $v_{200} = v_{\rm c,max} / (200 \, {\rm km \, s^{-1}})$ and $r_{200}$ is the radius encompassing a mass density 200 times the cosmological critical density. The second is based on the precipitation limit.  We assume that the ratio between $t_\text{cool}$ and $t_\text{ff}$ is maintained near $t_\text{cool} / t_\text{ff} = 10$ by approximate balance between radiative cooling and precipitation-regulated feedback, which tunes the electron density so that
\begin{figure}
  \centering
    \includegraphics[width=0.45\textwidth]{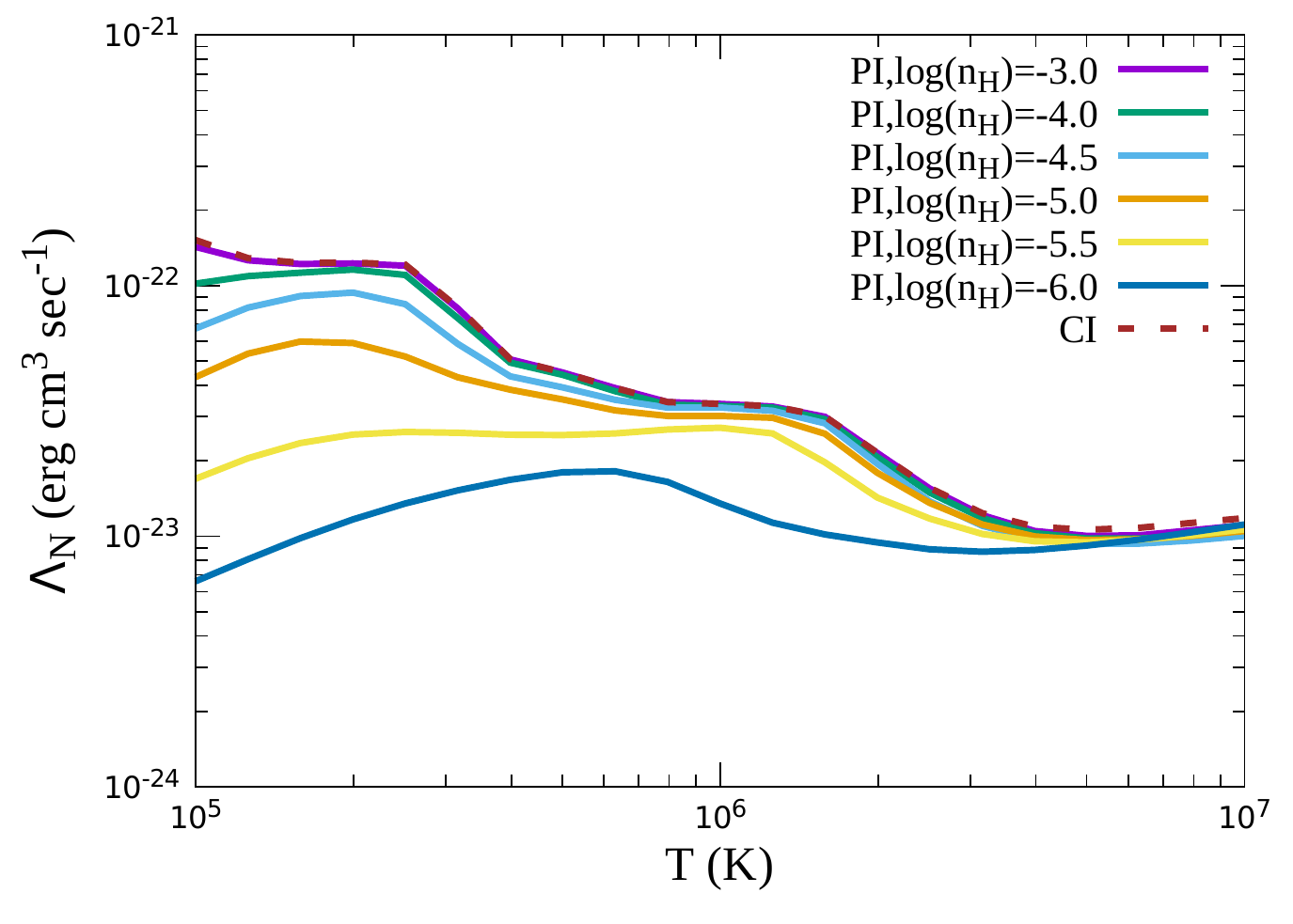}
    \caption{{Radiative cooling function $\Lambda_N (T)$ for gas {with metallicity 0.3 Z$_\odot$ and solar abundance ratios \citep{GASS10}} in pure collisional ionization (CI) equilibrium and with additional density-dependent photoionization by the \citet{HM12} extragalactic UV background {at redshift z=0.}}
    \label{cooling_curve}}
\end{figure}  
\begin{equation} \label{ne}
    n_{e,\text{pre}} (r) = \frac{3kT(r) }{10 \, \Lambda_\text{N} [ T (r) ]}
        \left( \frac{n_e n}{2n_{\rm H}^2} \right) 
        \frac{v_\text{c} (r) }{ \sqrt{2} \, r} 
        \; .
\end{equation}
Given this density profile (equation \ref{ne}) and a temperature profile $T(r)$, we can then write the precipitation limited entropy profile as
\begin{equation} \label{Kpre}
    K_\text{pre}=\left[ \frac {2 kT(r)} {\mu m_p v_\text{c}^2(r)} \right]^{1/3} 
        \left\{ \frac{10}{3} \left( \frac{2n_{\rm H}^2}{n_e n} \right)  
        \Lambda_\text{N} [ T(r)] \, r \right\}^{2/3} \; .
\end{equation}

We use CLOUDY \citep{Gary2017} to determine $\Lambda_{\rm N}(T)$, assuming a metallicity of $0.3 \, Z_\odot$, motivated by CGM absorption-line observations \citep{Pro2017}. Note that equation (\ref{Kpre}) differs from the corresponding equation in \citet{Voit2019}, where $K_{\rm pre}$ was approximated assuming that $T(r)/v_\text{c}^2(r)$ remains nearly constant and that $\Lambda_\text{N}(T) \approx \Lambda_\text{N}(2T_\phi)$, where $kT_\phi \equiv 0.5 \mu m_p v_\text{c}^2$.  \citet{Voit2019} then used the resulting entropy profile to obtain a more accurate temperature profile through integration of the hydrostatic equilibrium equation, given a boundary condition \begin{equation} \label{BoundaryTemp}
    k T(r_{200}) = 0.25 \, \mu m_p v_{\rm c,max}^2
\end{equation}
inspired by direct observations of galaxy clusters \citep[e.g.,][]{Ghirardini2019}.

\cite{Esmerian2020} have recently critiqued the approach of \cite{Voit2019}, because they find that $T(r) \propto v_c^2(r)$ is not a good approximation for the median CGM temperature near the virial radii of the simulated galaxies they analyze. Their precipitation limited CGM models instead rely on a custom temperature profile for each galaxy, derived from the simulations, to determine the entropy profile that gives $t_\text{cool} / t_\text{ff} = 10$ in hydrostatic equilibrium.  However, CGM temperature profiles are not generally available for real galaxies, requiring some sort of assumption to be made about them.  

Here we follow \cite{Voit2019} by applying the boundary condition in equation (\ref{BoundaryTemp}) but iteratively solve the hydrostatic equilibrium equation using a 4th-order Runge-Kutta scheme to determine the temperature profile at which $t_\text{cool} / t_\text{ff} = 10$, given our assumed entropy profile.  Figure \ref{profile} shows that the CGM temperature and density profiles we obtain differ by only $\approx8-10\%$ from the profiles of \cite{Voit2019}.  Consequently, the approach of \cite{Voit2019} is adequate for the assumed boundary condition, but is less accurate for the galaxies in \cite{Esmerian2020}, which have $k T(r_{200}) \ll 0.25 \, \mu m_p v_{\rm c,max}^2$.

We construct the entropy profile that goes into the integration by adding the entropy profiles in equations (\ref{Kbase}) and (\ref{Kpre}) to get 
\begin{equation}\label{KpNFW}
    K_\text{pNFW}(r) = K_\text{pre}(r) + K_\text{base} (r)  \; .
\end{equation}
The subscript indicates a precipitation-limited NFW model, as in \cite{Voit2019}. Using the fact that $n_e = ( kT / K_\text{pNFW})^{3/2}$, we can write the second term on the RHS of equation (\ref{HSE}) as
\begin{equation} \label{dndx}
 \frac{T}{n} \frac{dn}{dx} \: = \: \frac{T}{n_e} \frac{dn_e}{dx} 
        \: = \: \frac{3}{2}\frac{dT}{dx} - \frac{3}{2} \frac{T}{K_\text{pNFW}} \frac{dK_\text{pNFW}}{dx} 
\end{equation}
Using equation (\ref{dndx}), the hydrostatic equilibrium equation to be solved boils down to 

{\begin{equation}
\begin{split}
 \frac{dT}{dx} \: &= \: \frac{2}{5} 
    \left[ - \left( \frac{\mu m_H}{k} \right) \frac{d\phi}{dx} 
        + \frac{3}{2}\frac{T}{K_{pNFW}} \frac{d  K_{pNFW}}{d x} \right]  \, \\
    &= \frac{2}{5} 
    \left[ - \left( \frac{\mu m_H}{k} \right) \frac{d\phi}{dx} 
        + \frac{3}{2}\frac{T}{x} \frac{d \ln K_{pNFW}}{d\ln x} \right] .    
        \label{iter}
    \end{split}    
\end{equation}}
Equation (\ref{iter}) is the one that we iteratively solve to determine a temperature profile that gives $t_\text{cool} / t_\text{ff} = 10$ for $K_{\rm pNFW} = K_{\rm pre}(r) + K_{\rm base} (r)$, given the boundary condition $k T(r_{200}) = 0.25 \, \mu m_p v_{\rm c,max}^2$.  

Notably, the solution of equation (\ref{iter}) {for the temperature profile} is independent of the entropy profile's normalization.  Variation of the boundary condition significantly changes the temperature profile only in the outer region, where $T \le 10^6$ K.  Hence, the choice of boundary condition has only minor effects on the column densities of OVII and OVIII but can have considerably larger effects on the OVI column density. 
 
Our standard model for a galaxy like the Milky Way assumes $v_{\rm c,max} = 220 \, {\rm km \, s^{-1}}$, which corresponds to a halo mass of $2 \times 10^{12} \, M_\odot$.  It assumes that collisional ionization equilibrium (CI) in the CGM at the temperature $T(r)$ determines all the ionization fractions.  However, we also consider two modifications to the standard model, one in which photoionization (PI) by extragalactic background radiation alters the CGM ionization states and another in which log-normal fluctuations in gas temperature alter the CGM ionization states. 


\subsection{Effect of photoionization}

For photoionization, we use the Extra-galactic UltraViolet background radiation (UVB) model at redshift z=0 from \cite{HM12}.  This background radiation alters our CGM model because cooling functions with PI can differ from the pure CI case. Figure (\ref{cooling_curve}) shows how PI changes the CGM cooling function in gas of differing density. The differences from CI are small for $n_{\rm H} \gtrsim 10^{-4.5} \, {\rm cm^{-3}}$ but can be large in gas of lower density, particularly for $T < 10^6 \, {\rm K}$. {These results are consistent with the time independent cooling curves of CI and PI from previous studies \citep{Gnat_2017,Qu2018} in the presence of the same Extra-galactic UVB model.}



Our PI models therefore use a 2-dimensional interpolation scheme (in density and temperature) to the $\Lambda_{\rm N}$ curves in Figure (\ref{cooling_curve}) to iteratively solve for $n_e$ using equation (\ref{ne}).  We apply the outer temperature boundary condition to calculate the density there.  Then we advance inward using 4th-order Runge-Kutta integration for equation (\ref{iter}) as in the CI case.  Figure \ref{profile} shows the differences between the CI and PI model for our standard Milky Way-sized galaxy and also a low-mass galaxy ($v_{\rm c,max} = 90 \, {\rm km \, s^{-1}}$, $M_{\text{vir}}=10^{10}$ M$_\odot$). As the curves show, the differences are small for massive galaxies, in which $n_{\rm H} \gtrsim 10^{-4.5} {\rm cm^{-3}}$ at most radii, and are considerably larger for low-mass galaxies, in which $n_{\rm H} \lesssim 10^{-4.5} {\rm cm^{-3}}$ at most radii.

\subsection{Temperature Fluctuations}

We allow for the possibility of an inhomogeneous CGM by considering a log-normal temperature distribution around the median $T(r)$ at each radius.  The observed widths and centroid offsets of OVI absorption lines in the CGM suggest dynamical disturbances that may lead to temperature fluctuations. There can be several sources for these fluctuations. Outflows or high-entropy bubbles can lift low-entropy gas, and adiabatic cooling of these uplifted gas parcels as they maintain pressure balance gives rise to temperature and density fluctuations. Alternatively, turbulence-driven nonlinear oscillations of gravity waves in a gravitationally stratified medium can also lead temperature fluctutations as well as condensation resulting in multiphase gas. 

\cite{Faerman2017} considered an isothermal, two-phase model, one phase with temperature $\sim 1.5 \times 10^6$ K and another (designed to reproduce OVI measurements) at $\sim 3\times 10^5$ K. They assumed temperature fluctuations with a log-normal distribution around these two median temperatures.  They found that $\sigma_{\ln T} \thickapprox 0.3$ can account for the observed OVI column density in the context of their two-component CGM model. 

\citet{Voit2019} applied a similar fluctuation distribution around a single median gas temperature and showed that for the case of M$_{200}\geqslant$ 10$^{11.5}$ M$_\odot$, temperature fluctuations enhance the OVI column density by altering the ionization fraction at each radii. He also showed that each ion is generally present over a broader range in median temperature when temperature fluctuations are considered. Comparing with data of \citet{Werk2016}, he concluded that broad OVI absorption lines in the CGM of galaxies like the Milky Way are consistent with $\sigma_{\ln T} \thickapprox 0.7$. 

In this paper, we follow \citet{Voit2019} and consider how different values of $\sigma_{\ln T}$ around a single median temperature affect the predicted absorption-line column densities of highly ionized oxygen and estimate $\sigma_{\ln T}$ through comparisons with observations.

\section{Results for the Milky Way}
\label{Results}

\begin{figure*}
\centering
\includegraphics[width=1\linewidth]{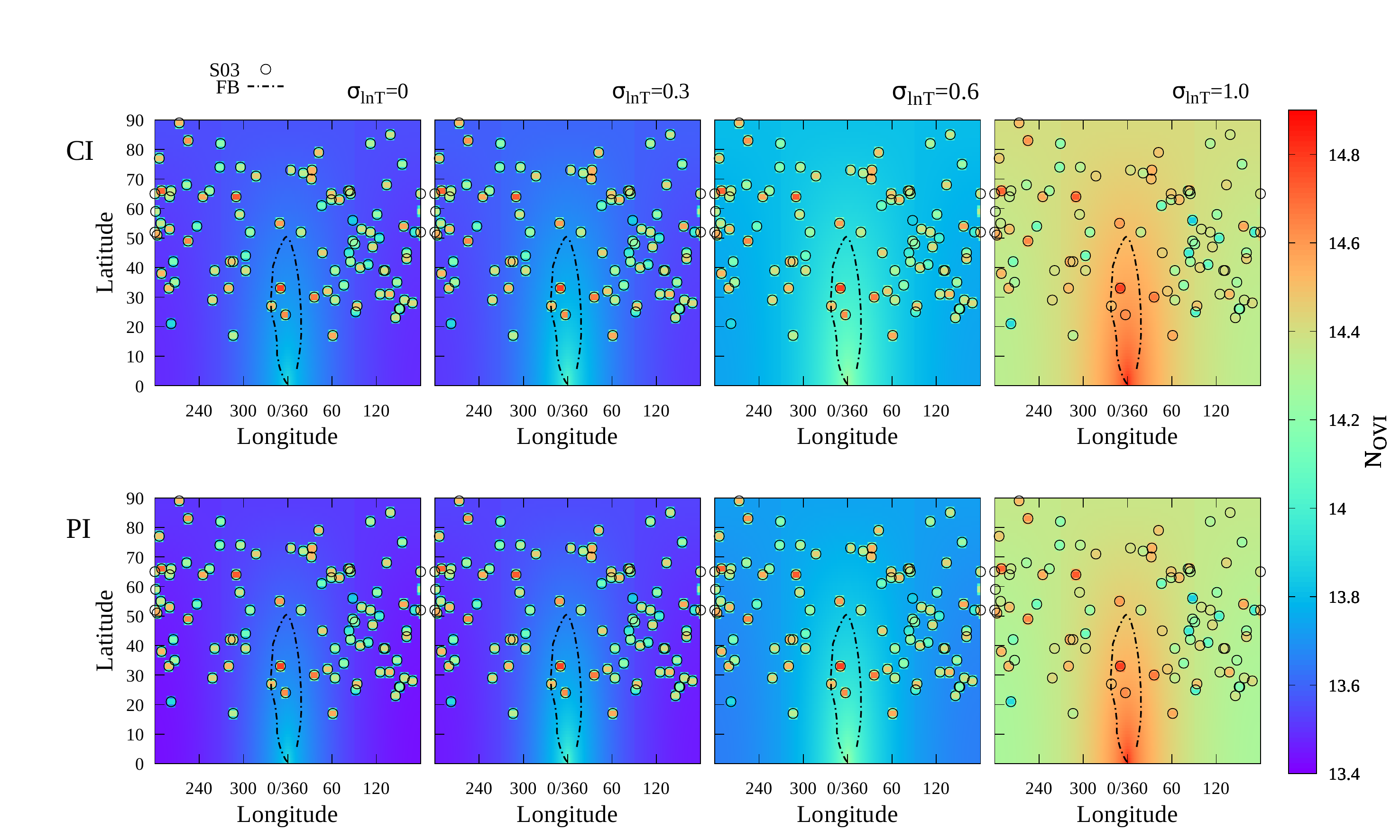}
\caption{Dependence of OVI on galactic latitude ($l$) and longitude ($b$) in the case of pure collisional ionization (CI, top row) and additional photoionization (PI, bottom row) by the UV background, given different $\sigma_{\ln T}$. {The observed value of N$_{\rm OVI}$ by \citet{Savage2003} from {\it Far-Ultraviolet Spectroscopic Explorer (FUSE)} spectra of 100 extragalactic objects and two distant halo stars have been shown by circles with appropriate colours. The observation along negative latitudes are shown as positive latitudes. We have excluded data points that give only upper-limits.} The gamma-ray edge of the Fermi bubble is superposed on this plot, denoted by dashed lines. }
\label{NOVI}
\end{figure*}

\begin{figure*}
 \centering
    \includegraphics[width=1.0\textwidth]{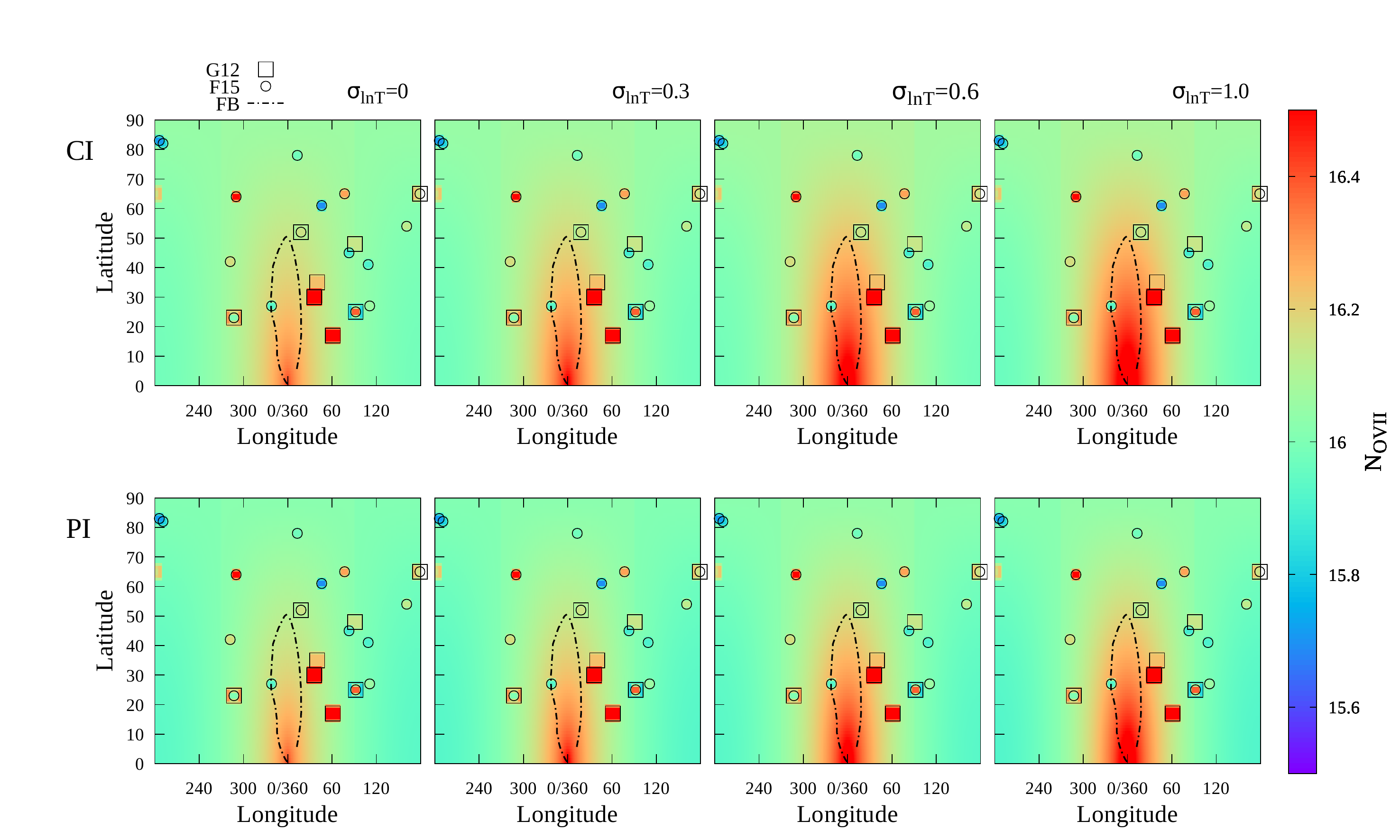}
    \caption{Dependence of OVII on galactic latitude ($l$) and longitude ($b$) in the case of pure collisional ionization (CI, top row) and additional photoionization (PI, bottom row) by the UV background, given different $\sigma_{\ln T}$. {The observed values of N$_{\rm OVII}$ by \citet{Gupta2012} and \citealt{Fang2015} have been shown by squares and circles respectively with appropriate colours. The observation along negative latitudes are shown as positive latitudes. The observational uncertainties in the observations of \citet{Gupta2012} are typically a factor of $1.5-3$ whereas same is larger (factor of $1.6$ to even $500$) in the case of \citet{Fang2015} as their saturation corrections are more uncertain. We plotted 16 lines of sight out of 33 from \citet{Fang2015} as their observed OVII column density distribution peaks around (log OVII) $16.0$, with half the data points (16 out of 33) in the range of $15.5-16.5$ (the right panel of fig 12 from their paper).} The gamma-ray edge of the Fermi bubble is superposed on this plot, denoted by dashed lines.
    \label{NOVII}}
  \end{figure*}
  
\begin{figure*}
\centering
\includegraphics[width=1\linewidth]{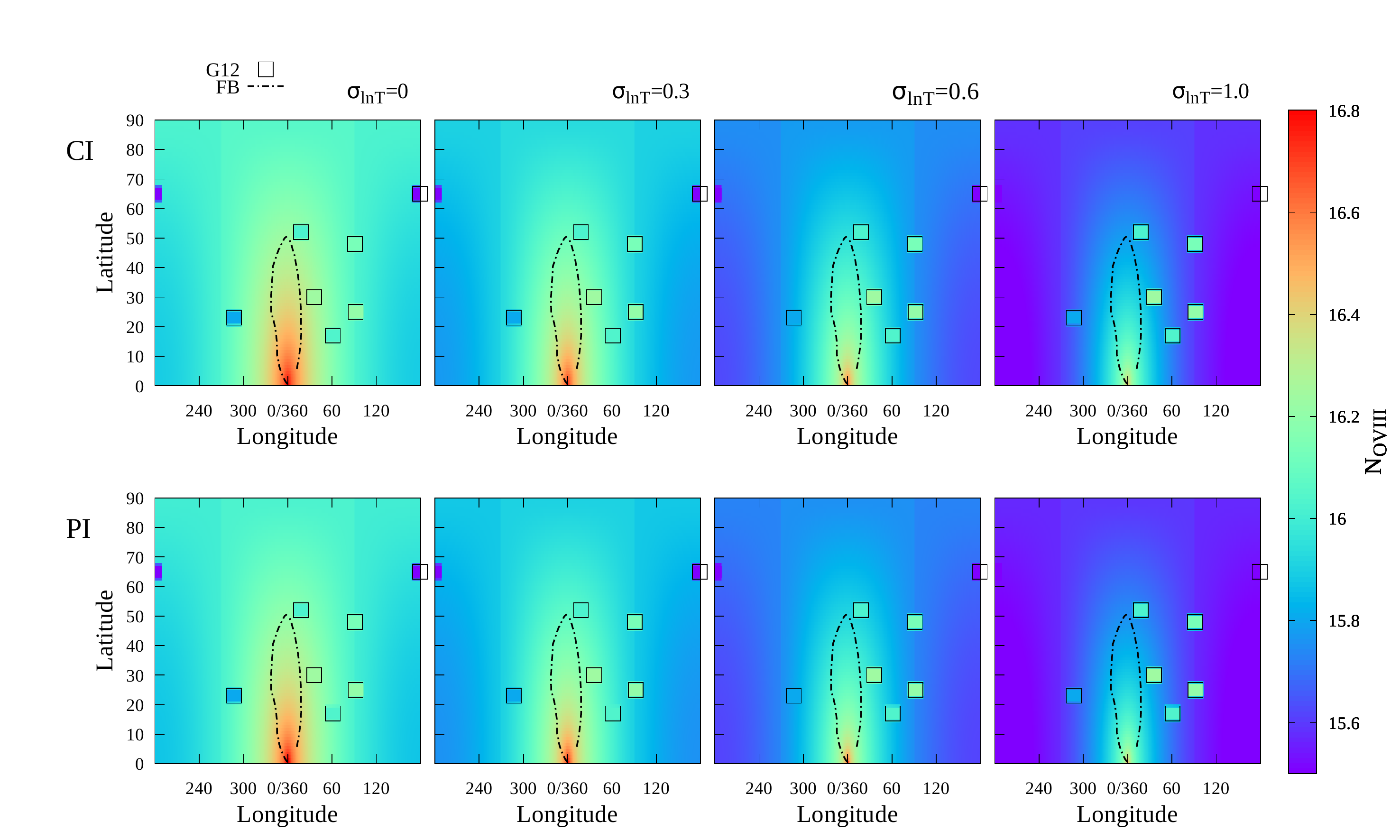}
\caption{Dependence of OVIII on galactic latitude ($l$) and longitude ($b$) in the case of pure collisional ionization (CI, top row) and additional photoionization (PI, bottom row) by the UV background, given different $\sigma_{\ln T}$. {The derived values of N$_{\rm OVIII}$ (assuming no saturation) from observed equivalent widths  by \citet{Gupta2012} have been shown by squares with appropriate colours. The observation along negative latitudes are shown as positive latitudes.} The gamma-ray edge of the Fermi bubble is superposed on this plot, denoted by dashed lines. }
\label{NOVIII}
\end{figure*}

Here we use the ingredients described so far to calculate the absorption-line column densities of highly-ionized oxygen along different lines of sight through the Milky Way's CGM that all start at the solar system.  Some previous calculations have considered CGM column density as a function of impact parameter through the CGM and have chosen a particular value of impact parameter for comparisons with observations. \cite{Qu2018} have taken $0.03$ $\text{R}_{\text{vir}}$ ($\approx 8 \, {\rm kpc}$) as their chosen impact parameter, while comparing with column density data for $b>60^\circ$, because the column density at this impact parameter would be twice the value of column density seen at $b=90^\circ$ (the distance of the Galactic centre from solar system being $\approx 8$ kpc). 
However, while comparing the predicted column density at this impact parameter, they used the median value of the whole data set, which includes data from low Galactic latitudes.
{We note that \cite{Faerman2017} and \cite{Faerman2019} have shown the variation of column density  with respect to the angle from Galactic Centre for an observer at solar position. However, for comparison with observations they used the value of column density which is half of its total value up to r$_{\rm CGM}$.}

\cite{MB2013} have analyzed a sample of 29 sightlines with absorption line observations of OVII and OVIII. They considered two isothermal models, a spherical model and a flattened model, and attempted to fit the observed column densities by transforming the galactocentric density profile to a coordinate system centered on the Sun.
 
 \begin{figure}
  \centering
    \includegraphics[width=0.5\textwidth]{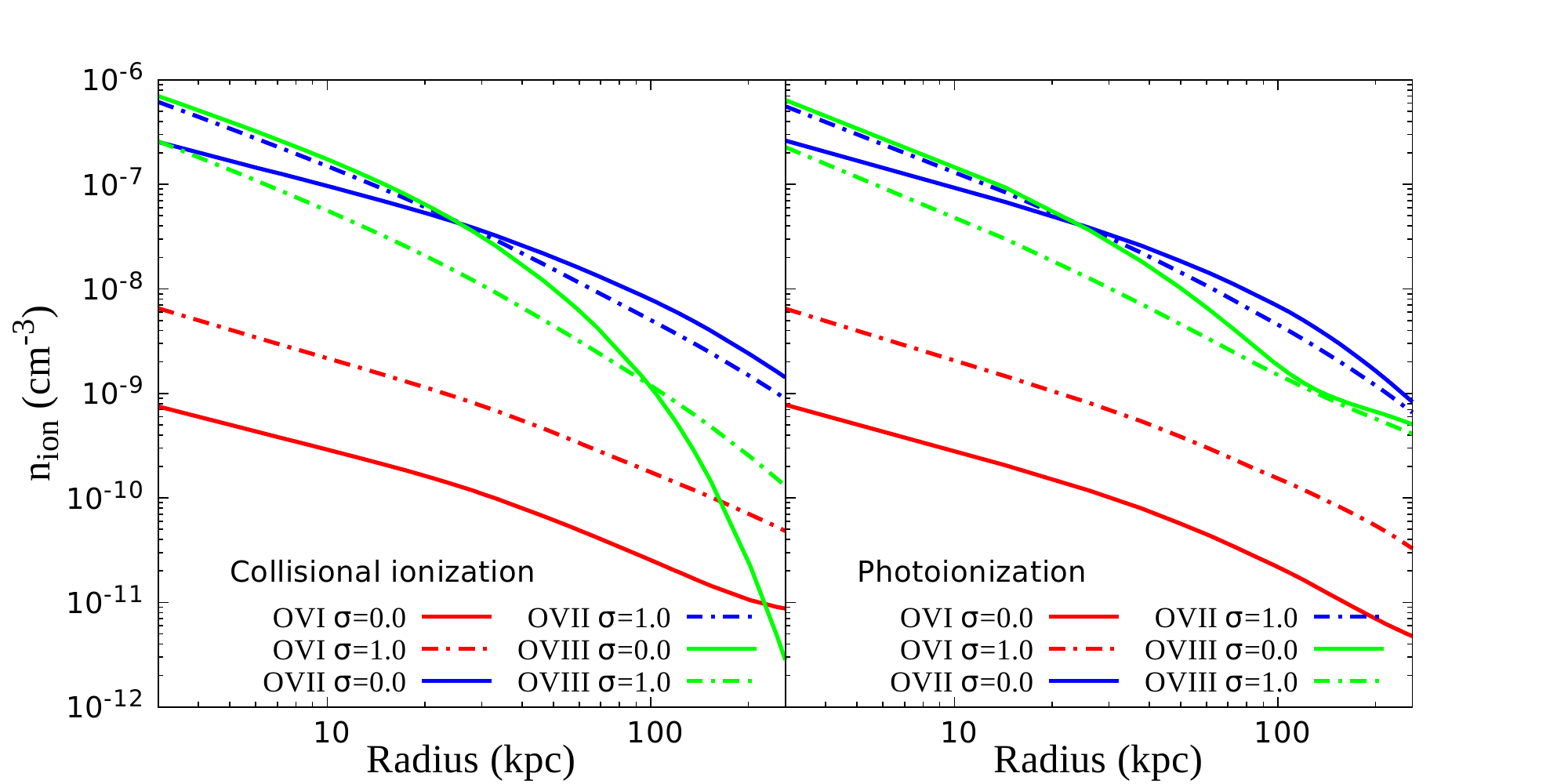}
    \caption{Radial density profiles of OVI, OVII, and OVIII for M$_{\rm vir} = 2\times10^{12} \, M_\odot$.}
    \label{radialprofile}
  \end{figure}

Since the observations with which we wish to compare our results involve different lines of sight through the CGM as seen from our location, we also calculate predicted column densities from the Sun's location as a function of Galactic latitude ($l$) and longitude ($b$). 
The predicted  {OVI,} OVII, and OVIII column densities are presented in figures {(\ref{NOVI})},  (\ref{NOVII}) and (\ref{NOVIII}) respectively.  In each figure, the upper rows show the case of only collisional ionization (CI) and the lower rows include photoionization (PI) by the extragalactic UVB. The leftmost columns of both figures show results when temperature fluctuations are not included ($\sigma_{\ln T}=0$) and the columns to the right show the effects of increasing $\sigma_{\ln T}$ to 0.3, 0.6, and 1.0. 

The figures show substantial variations in the column densities of OVI, OVII, and OVIII with both longitude $l$ and latitude $b$.  Temperature fluctuations also have significant effects. For example, the range of OVI column density predicted by CI models changes from  $13.5 \le \log N_{\rm OVI} \le 13.8$ for $\sigma_{\ln T}=0$ to $14.4 \le \log N_{\rm OVI} \le 14.8$ for $\sigma_{\ln T}=1.0$.  OVII column density ranges between $15.9 \le \log N_{\rm OVII} \le 16.4$ for $\sigma_{\ln T}=0$, whereas for $\sigma_{\ln T}=1.0$, the range changes to $15.9 \le \log N_{\rm OVII} \le 16.9$. For OVIII column density, the corresponding ranges change from $16.0 \le \log N_{\rm OVIII} \le 16.8$ for $\sigma_{\ln T}=0$ to $15.5 \le \log N_{\rm OVIII} \le 16.8$ for $\sigma_{\ln T}=1.0$.  In other words, temperature fluctuations {\it decrease} the total OVIII ion density and {\it increase} the total OVI and OVII column densities, with most of the OVIII change coming from the inner CGM.  Figure (\ref{radialprofile}) shows this effect by plotting the density profiles of different ions for models with $\sigma_{\ln T}=0$ and $1.0$. 
 
Figure (\ref{radialprofile}) also shows differences between the purely collisional (CI) and photoionization (PI) cases.  Around a galaxy like the Milky Way, the main effect of photoionization is on the OVIII ion, especially at outer radii in the case with no temperature fluctuations.  (Section \ref{LowMassGalaxies} shows that UVB photoionization can have considerably greater effects on precipitation-limited CGM models for low-mass galaxies.)  The effects on OVIII arise because the ionization potential of OVII is much greater than that of OVI.  Therefore, collisionally ionized gas in the temperature range we are concerned with here (a few times $10^5$ K to $\sim 2 \times 10^6$ K), has a relatively small OVIII number density ($n_\text{OVIII}$) at large radii. But the inclusion of UVB photons (which can have energies of a few hundred eV) can increase the OVIII number density. This effect becomes particularly significant when the density is low, which happens at the outer radii. {It was also pointed out by \cite{Faerman2019} in their isentropic model. However, here we show that} the effect of temperature fluctuations can compensate for the effects of photoionization. The resulting increase in OVIII number density in the case of PI (green solid line, without temperature fluctuation) is seen in Figure \ref{radialprofile}, when the corresponding two panels are compared. But when temperature fluctuation is included (green dotted lines), the radial profiles of OVIII number density for PI and CI cases are similar at large radii. 

 

\begin{figure*}
  \centering
  \includegraphics[width=1.\linewidth]{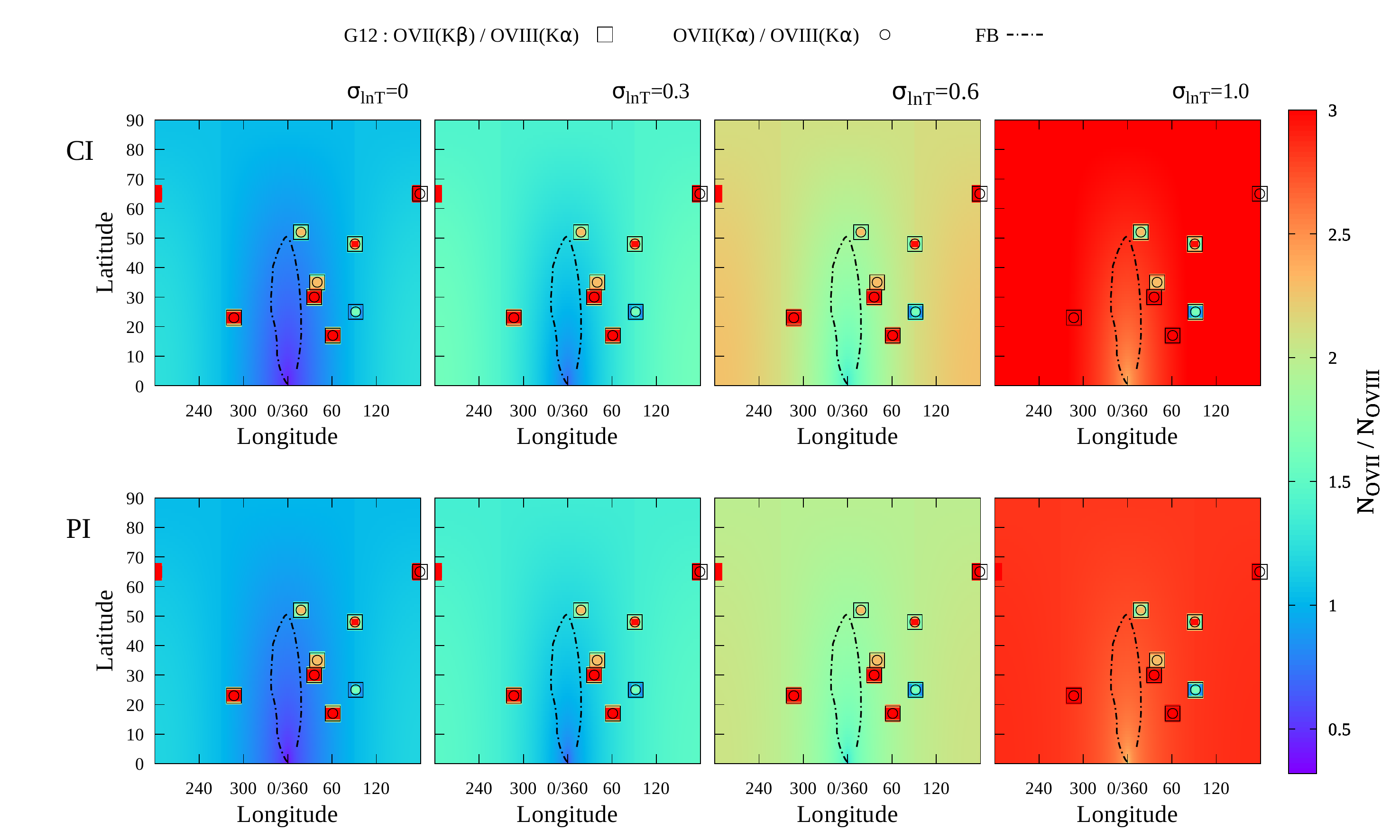}
    \caption{Dependence of the column-density ratio $N_\text{OVII} / N_\text{OVIII}$ on galactic latitude ($l$) and longitude ($b$) in the case of pure collisional ionization (CI, top row) and additional photoionization (PI, bottom row) by the UV background, given different $\sigma_{\ln T}$. {The derived values of ratio between N$_{\rm OVII}$ and N$_{\rm OVIII}$ from observation of \citet{Gupta2012} have been shown by squares and circles with appropriate colours. The observation along negetive latitudes are shown as positive latitudes.} The gamma-ray edge of the Fermi bubble is superposed on this plot, denoted by dashed lines.}
    \label{ratio}
      \end{figure*} 


Going back to Figures (\ref{NOVII}) and (\ref{NOVIII}), we note that column densities can be rather large for $l \sim0, b\le 50^\circ$. However, sight lines passing close to the Galactic centre are not well suited for comparison with observations, owing to the Fermi bubbles in this range of latitudes and longitudes. Dot-dashed lines outline the gamma-ray edge of the Fermi bubbles, following \cite{Su2010}, in order to exclude these sightlines from our present consideration. 

Figure (\ref{ratio}) shows the predicted ratio of OVII to OVIII column density as a function of Galactic latitude and longitude for both CI (upper panels) and PI (lower panels) and different values of $\sigma_{\ln T}$. 
Temperature fluctuations significantly decrease N$_{\rm OVIII}$ compared to N$_{\rm OVII}$, thereby increasing the value of the ratio, as can be seen by looking from left to right in figure (\ref{ratio}). The predicted N$_{\rm OVII}$/N$_{\rm OVIII}$ ratios span a large range, from $\sim 0.5$ for $\sigma_{\ln T}=0$ (along sightlines toward the Galactic centre) to $\sim 3.0$ for $\sigma_{\ln T}=1.0$ (along  sightlines away from the Galactic centre). Comparing the CI and PI cases for a given value of $\sigma_{\ln T}$, shows that the N$_{\rm OVII}$/N$_{\rm OVIII}$ ratio slightly decreases from the CI to PI case, because of the PI increase in OVIII density explained above. 

\section{COMPARISON WITH {UV and} X-RAY OBSERVATIONS}
\label{OBS}

This section compares our model predictions for the Milky Way's CGM with {UV and} X-ray observations. { \cite{Savage2003} used the observation of {\it FUSE} far-UV spectra of 100 extragalactic objects and
two halo stars to measure the OVI column density along different lines of sight through the Milky Way. We compare our model predicted values with these observations and superpose them on our OVI map of the Milky Way (Figure \ref{NOVI}). Note that we have not plotted data points that give only upper limits and the observation along negative latitudes are shown as positive latitudes. 
{The observed values of log(N$_{\rm OVI}$) are mostly large ($14.8-14.9$) in the vicinity of Fermi bubble region whereas at larger longitudes their values are mostly smaller ($14.4-14.6$). This  trend along with the observational range of log(N$_{\rm OVI}$) agrees with the prediction of our model for $\sigma_{\ln T}=1.0$.}}

{The emission-line intensities of OVII and OVIII are similar to the predictions of \citet{Voit2019}, which presents a comparison with the data on X-ray emission. So, we here focus only on absorption data in order to compare with our model.} Table \ref{observation} summarises the reported measurements of OVII and OVIII lines in X-ray absorption spectra. However, some explanations are in order regarding the values in the table. 

\cite{Gupta2012} have reported measurements of $N_{\rm OVII}$ and $N_{\rm OVIII}$ with {\it Chandra} along eight sight lines. Their observed values of log(N$_{\rm{OVII}}$) range from 15.82 to {16.7} with a weighted sample mean of 16.19. \cite{MB2013} have studied OVII lines {toward} 26 AGNs along with one detection of {an OVIII line with a ratio N$_{\rm{OVII}}$/N$_{\rm{OVIII}}$=$0.7 \pm 0.2$.}  \cite{Fang2015} have also measured OVII column density from data on 43 AGNs, including the sample of \cite{MB2013}, observed with {\it XMM-Newton}. They found that most of their OVII lines had column densities centered around 10$^{16}$ cm$^{-2}$ with a wide range, $10^{15.5}-10^{16.5}$ cm$^{-2}$. {The observed values of N$_{\rm OVII}$  by \citet{Gupta2012} and \citet{Fang2015}  have been shown in Figure \ref{NOVII} by circles and squares respectively with appropriate colours. We plotted 16 lines of sight out of 33 from \citet{Fang2015} as their observed OVII column density distribution peaks around (log OVII) $16.0$, with half the data points (16 out of 33) in the range of $15.5-16.5$ (the right panel of fig 12 from their paper). The derived values of N$_{\rm OVIII}$ (assuming no saturation) from observed equivalent widths by \citet{Gupta2012} have been shown in Figure \ref{NOVIII} by squares with appropriate colours. Note that the observations along negative latitudes are marked as positive latitudes.} 

{However, \cite{Faerman2017} increased \citet{Gupta2012}'s observed values of OVII and OVIII equivalent widths (EW) by 30\% in order to correct for systematic error and re-calculated the median as $16.34 (16.25 - 16.46)$. For 10 objects in \citet{Fang2015}'s sample, they could derive only upper limits and calculated the range of the OVII column density by ignoring those sight lines.  \cite{Faerman2017} took into account the upper limits for the non-detections and added those upper limits to the group of detected absorption lines. This addition reduced the median column density of the \cite{Fang2015} sample to $16.15 (16 - 16.3)$. The third and fourth columns of table \ref{observation} show both the original measurements and the revised values by \cite{Faerman2017} of $N_{\rm OVII}$ and of $N_{\rm OVIII}$ respectively.}

\cite{Fang2015} did not report any OVIII observations, and \cite{Gupta2012} gave only equivalent widths for OVIII absorption. \cite{Faerman2017} calculated the corresponding column densities (and the median value thereof) assuming that the OVII and OVIII lines arise in similar physical conditions. These values are shown in the 4th column of table \ref{observation}.  \cite{Faerman2017} then derived a median value of $4.0$ for the ratios of column densities along individual sight lines from \cite{Gupta2012}. They used this value of the ratio to estimate the OVIII column densities in \cite{Fang2015} sample, which are shown in the second row of the 4th column.

In the second row of 5th column, we show the median ratio {estimated} by \cite{Faerman2017}. Note that this is the median of ratios of corrected column densities along each of the eight sight lines from \cite{Gupta2012}. Also note that the ratio of the medians ($2.1$) is different from the median of ratios ($4.0$). The ranges of values shown for \cite{Faerman2017} refer to a 1-$\sigma$ uncertainty around the median, although it is the {standard deviation around} the mean which they applied to the median.

We can, however, obtain the ratio of column densities of OVII and OVIII from the observations by \cite{Gupta2012} in a slightly different manner, which might be physically more meaningful.  Along 6 out of 8 lines of sight in \cite{Gupta2012}, the EW of {\rm{OVII} K$_\beta$ and \rm{OVIII} K$_\alpha$} are essentially the same, given the observational uncertainities. The lines also have similar wavelengths. Therefore, the optical depths of those lines are similar, which implies that they experience approximately the same amount of saturation. If both the optical depths and the wavelengths are similar, then the ratio of column densities is approximately equal to the inverse of the ratio of oscillator strengths, {\it i.e.}  $N_{\rm{OVII}}/N_{\rm{OVIII}}\approx2.8$. However, one should consider the large uncertainties in the EW ratios.  For the 6 lines of sight with detected {\rm{OVII} K}$_\beta$, we get a weighted mean of the ratio of EWs of $0.7\pm0.2$.  Dividing this mean EW ratio by the ratio of oscillator strengths gives $N_{\rm{OVII}}/N_{\rm{OVIII}}\approx2.0 \pm 0.6 $. {The ratios along each sightline calculated from {\rm{OVII} K$_\beta$ and \rm{OVIII} K$_\alpha$} lines are shown in Figure \ref{ratio} by squares with appropriate colours.} If we consider the {\rm{OVII} K$_\beta$ and \rm{OVIII} K$_\alpha$} instead, we  find the weighted mean of EW ratios to be $2.3\pm0.2$ which gives $N_{\rm{OVII}}/N_{\rm{OVIII}}\approx1.1 \pm 0.1 $. This is, however, only a lower limit because the actual ratio is greater, if the lines are saturated. A mean saturation correction can be derived from the OVII EW ratios EW(K$_\beta$)/EW(K$_\alpha$) measured by \cite{Gupta2012}, which have a weighted mean of  $0.43 \pm 0.06$. Comparing with the expected ratio of $0.156$ for the optically thin case, this implies a saturation correction factor of $2.7 \pm 0.4$. If we apply this correction to the ratio of column densities we get $N_{\rm{OVII}}/N_{\rm{OVIII}}\approx3.0 \pm 0.5$. {The ratios along each sightlines calculated from {\rm{OVII} K$_\alpha$ and \rm{OVIII} K$_\alpha$} lines are shown in Figure \ref{ratio} by circles with appropriate colours. Please note that the observation along negetive latitudes are shown as positive latitudes. } The ratio of these two estimates can also be averaged to give the $N_{\rm{OVII}}/N_{\rm{OVIII}}\approx2.6\pm0.4$. This value is quoted in the 1st row of the 5th column of table \ref{observation}. 


\begin{table*}
    \centering
	\caption{Summary of the observations} 
	\begin{tabular}{|| c c c c c ||} 
 \hline
 \label{observation}
Data set &  & log(N$_{\rm OVII}$) & log(N$_{\rm OVIII}$) & Ratio (N$_{\rm OVII}$/N$_{\rm OVIII}$)  \\ [0.7ex] 
 \hline\hline
 \multirow{2}{*}{\cite{Gupta2012}} & original & $16.19 \, (15.82-16.5)$ & -- & $2.6(2.2-3.0)$  \\ & \cite{Faerman2017} & 16.34 \, (16.25-16.46) & 16.0 \, (15.88-16.11) & 4.0 (2.8-5.62) \\ 
 \hline
 \multirow{2}{*}{\cite{Fang2015}} & original & $16.0 \, (15.5-16.5) ^*$ & -- & $0.7 (0.5-0.9) ^{**}$\\ & \cite{Faerman2017} & 16.15 \, (16.0-16.3) & 15.5 \,(15.3-15.7)  & -- \\
 \hline
 \end{tabular}
 \vspace{1ex}

     {\raggedright *Reported previously by \cite{MB2013} for smaller sample size. \par}

     {\raggedright **Reported ratio by \cite{MB2013} for one OVIII observation. \par}
\end{table*}

Comparing our models with the observational data from table \ref{observation} shows that the models with greater temperature fluctuations are in better agreement with the observations.  It can be seen that log(N$_{\rm OVIII}$) in our model ranges from $15.5$ to $15.7$ for $\sigma_{\ln T}=1.0$, and log(N$_{\rm OVII}$), ranges between $16.0$ to $16.2$ for the same value of $\sigma_{\ln T}=1.0$, for both CI and PI models, excluding the Fermi Bubble region. 
It is clear from Figures \ref{NOVII} and \ref{NOVIII} that there is not much difference in the predicted column densities of OVII and OVIII between CI and PI. The predicted ratio of N$_{\rm OVII}$ to N$_{\rm OVIII}$ for $\sigma_{\ln T}=1.0$ {is between $2.8$ and $3.0$ and is therefore} within the 1$\sigma$ uncertainty range around the median found by \cite{Faerman2017}, for both CI and PI. However, the model predictions for $\sigma_{\ln T}=0.6$ ($1.7$--$2.5$) are a better match to the ratio of the median column densities of OVII and OVIII.  Although, if we consider the  ratio of column densities we have re-estimated in this paper from the observations by \cite{Gupta2012}, we find that $\sigma_{\ln T}=0.6$ is consistent with the observations of \rm{OVII} K$_\beta$ and \rm{OVIII} K$_\alpha$ lines, whereas the ratio is more consistent with our model with  $\sigma_{\ln T}=1.0$ for observations of \rm{OVII} K$_\alpha$ and \rm{OVIII} K$_\alpha$ lines. The weighted mean of these two ratios $2.6\pm0.4$ is consistent with the range of $\sigma_{\ln T} \approx 0.6-1.0$. Therefore, the observed ratios favor values of $\sigma_{\ln T} \approx 0.6-1.0$ in the context of our precipitation limited CGM model. 

One should also consider the individual values of N$_{\rm OVII}$ and N$_{\rm OVIII}$. If we compare our model values of N$_{\rm OVIII}$ with \cite{Gupta2012} observations, then $\sigma_{\ln T}=0.6$ can provide a good match. {The observed values of N$_{\rm OVIII}$ do not show the angular dependence predicted by the model. However, it is worth noting that one data point at high latitude and longitude (l$\sim180$ b$\sim65$) shows a lower log(N$_{\rm OVIII}$) than other 6 data points, and is in general agreement with the model predictions. But more observational data points, preferably with higher precision are required in order to draw any definite conclusion about the angular dependence of N$_{\rm OVIII}$.} The predicted range of log(N$_{\rm OVIII}$) $\approx 15.5\hbox{--}15.7$ for $\sigma_{\ln T}=1.0$  matches the range obtained by \cite{Faerman2017} for \cite{Fang2015} using the ratio of column densities from \cite{Gupta2012}. For log(N$_{\rm OVII}$), the predicted column density for {for all values of} $\sigma_{\ln T}$
(excluding the Fermi Bubble region) is in the range $\approx 16.0\hbox{--}16.3$, which falls within the originally reported values of  \cite{Gupta2012} and {original as well as re-estimated values  of \cite{Fang2015} by \cite{Faerman2017}.} {Approximately 50 percent of the data points (includes \cite{Gupta2012} and \cite{Fang2015} both) are in good agreement with the angular variation predicted by our model.} {However, in Figure \ref{NOVII}, we find deviations of the observed values from the predicted ones mainly near the Fermi Bubble {and some at large longitude directions as well}. Note that, the uncertainties in the observations of N$_{\rm OVII}$ are large; factor of $1.5-3$ in case of \cite{Gupta2012} and even larger (factor of $1.6-500$) in the case of \cite{Fang2015}. However it is difficult to rule out any model solely on the basis of N$_{\rm OVII}$ because there is no significant change in N$_{\rm OVII}$ with $\sigma_{\ln T}$.}  

We can also compare the N$_{\text{OVII}}$ data for different values of $l$ and $b$ from \cite{Fang2015} with the computed values for $\sigma_{\ln T}=1.0$ from our model. 
We calculate the Pearson correlation coefficient between the model and the observed values and find the correlation to be $r \approx 0.2$. As mentioned earlier, the sight lines around $l \sim 0.0$ may not be suitable  for comparison because of contamination from the Fermi Bubble, whose boundary is marked in {Figure (\ref{NOVII}) and Figure (\ref{NOVIII})}. Performing a t-test for the data (for $N=33$ data points) yields a value of
\begin{equation}
 t=r \times \sqrt{ \left( \frac{N-2}{1-r^2} \right)} 
 \: \thickapprox \, 1.08
    \; .
\end{equation} 
This t-statistic is distributed in the null case of no correlation like a Student's t distribution with $\nu = N-2 = 31$ degrees of freedom, whose two-sided significance level is given by $1-A(t|\nu) = 0.28$.
However, we note that the highest column density observations are mainly concentrated in three zones of $l,b$.  Region 1 ($l \le 30, l\ge 350, b\le 50$) coincides with the direction of the Fermi Bubble. Region 2 ($260 \le l \le 300, b\le 60$) points towards the high-temperature zone of the Local Bubble. Therefore observations of sight lines in these two regions are likely to be contaminated by foreground rather than CGM. However, column densities are also high in Region 3  ($30 \le l \le 90, b\le 60$) and sight lines in these regions would provide the best diagnostic for CGM. 
If these sight lines are removed, and if we further focus on the sight lines that produce the column density $\le 10^{16}$ cm$^{-2}$, then the correlation coefficient increases to $\sim 0.3$ and the probability for null case decreases to $\sim 0.26$.  {But for both the above cases, the significance for rejecting null hypothesis is low because we have only a small number of observations to compare with. However if we compare larger N$_{\text{OVI}}$ data set for different values of $l$ and $b$ from \cite{Savage2003} with the computed values for $\sigma_{\ln T}=1.0$ from our model, we get correlation coefficient to be $0.35$, t value to be $3.54$ and probability for null case decreases to $\sim 10^{-4}$. Hence, larger observational data set improves in the significance for rejecting null hypothesis.}

In summary, the range of $\sigma_{\ln T}$ that best agrees with all the observational quantities is $\approx 0.6$--1.0. {The uncertainties in the data, However, are large, which makes it difficult to find a point-by-point match with the model. We note that the comparison shown by \cite{MB2013} for an off-centered model of the Milky Way CGM does imply an underlying angular variation in the column density data. 
More data with less uncertainties will definitely help to constrain the precipitation model in the future. 
}
  

\section{Discussion}
\label{Discussion}

\subsection{CGM in galaxies like the Milky Way} \label{OtherGalaxies}

\begin{figure}
  \centering
    \includegraphics[width=0.5\textwidth]{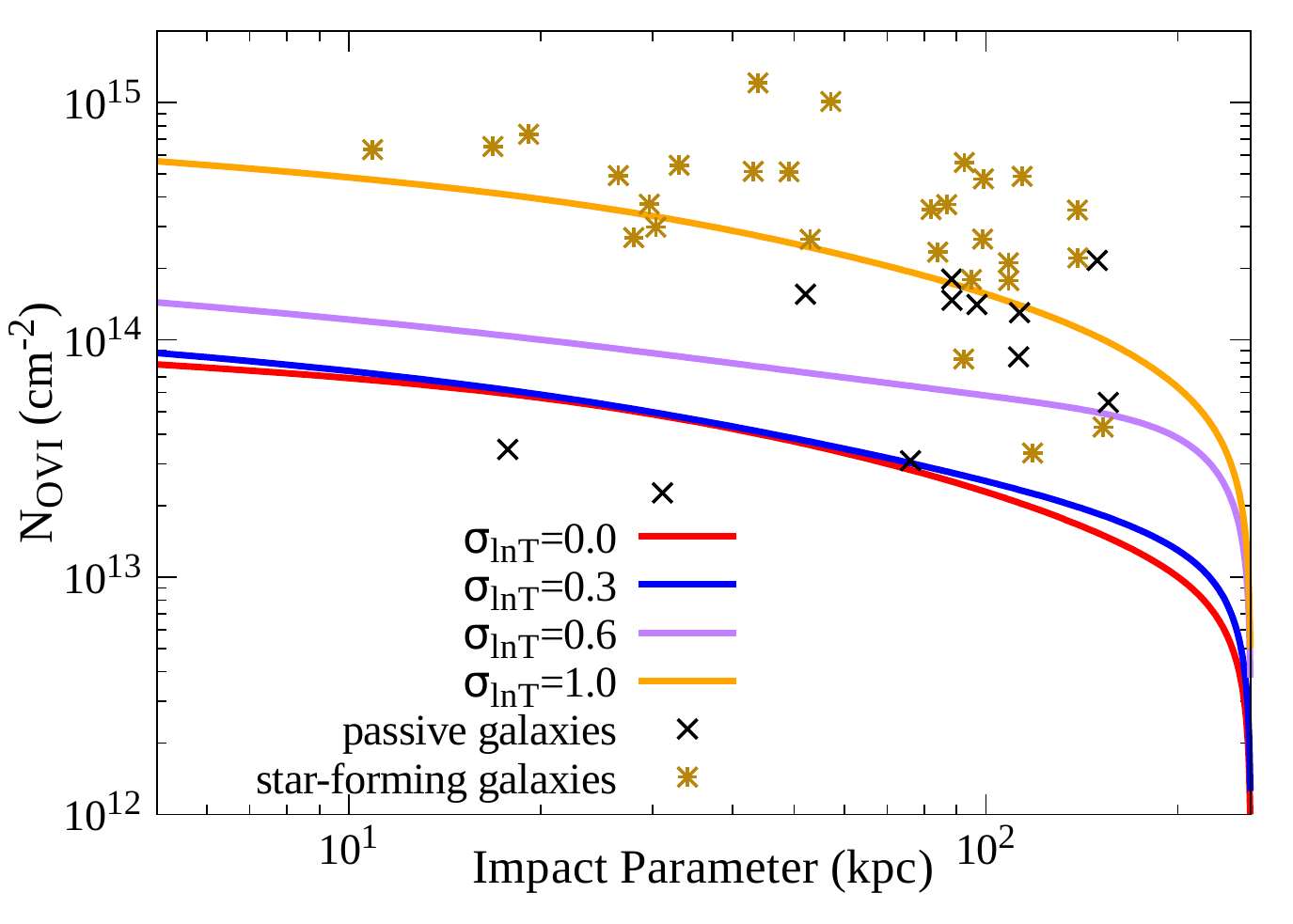}
    \caption{Variation of OVI column density with sigma and comparison with observation of star forming and passive galaxies from \citealt{Tumlinson2011}.}
    \label{SFR}
  \end{figure}

Encouraged by the apparent agreement between our precipitation models with $\sigma_{\ln T} \approx 0.6$--1.0 and OVII and OVIII column density observations for the Milky Way CGM, we now look towards extragalactic observations. Figure \ref{SFR} compares the calculated column density of OVI from our model with OVI observations of a set of star-forming and passive galaxies by \citep{Tumlinson2011}. It can be seen that large-amplitude temperature fluctuations {are required to fit the star-forming galaxies with the model}.  In the figure, log-normal fluctuations with $\sigma_{\rm{lnT}} \approx 1.0$ are needed to obtain the OVI column densities observed around star-forming galaxies, while smaller values of $\sigma_{\rm{lnT}}$ suffice for passive ones. This finding implies that star formation is somehow related to temperature fluctuations in the CGM.  

We note that the values of $\sigma_{\ln T}$ inferred here from OVI absorption are somewhat larger than reported in \cite{Voit2019}. This finding may arise from the following causes. Firstly, the default CGM metallicity in \cite{Voit2019} was solar, in contrast to $0.3$ Z$_\odot$ considered here, and N$_{\rm OVI}$ in precipitation-limited CGM models approximately scales as (Z/Z$_\odot$)$^{0.3} ${(see Eq. 17 of \citealt{Voit2019})}.
Secondly, {uncertainity in the virial radius and mass of  MW-type halos along with that in the position of the accretion shock leading to uncertainty in the CGM outer radius (r$_{\rm vir}$ to 1.3 r$_{\rm vir}$)} can introduce differences of $\sim 10$\% in predicted column densities. Thirdly, differences in the rates assumed for different atomic processes can lead to different predictions for the ionization fractions.  For example, the collisional ionization model of \cite{SD1993}, used by \cite{Voit2019}, has {different} dielectronic recombination rate coefficients {as well as solar abundances} from the ones {used here by} CLOUDY \citep{Gary2017} leading to {a} substantial differences in the OVI ionization fraction {(e.g., $6\times10^{-3}$ using CIE model of \cite{SD1993} whereas $4.6\times10^{-3}$ using CLOUDY at $10^6$~K)}.One must therefore be cautious {about these uncertainities while} drawing conclusions on the physical state of the CGM that depend on those {values}. 

\subsection{CGM in low mass galaxies} \label{LowMassGalaxies}

{The results we have presented so far indicate that photoionization does not play a significant role in determining the column densities of highly-ionized oxygen in the CGM of galaxies like the Milky Way.  However, precipitation-limited CGM models of lower mass galaxies have considerably lower densities, in which photoionization becomes more important.  Here we present CGM models of low-mass halos (M$_{\text{halo}}<2 \times 10^{11}$M$_\odot$) to illustrate how photoionization affects our CGM models for those galaxies.} {The entropy, density and temperature profiles for low mass galaxies are shown in figure \ref{profile}.}

\begin{figure*}
  \includegraphics[width=0.45\linewidth]{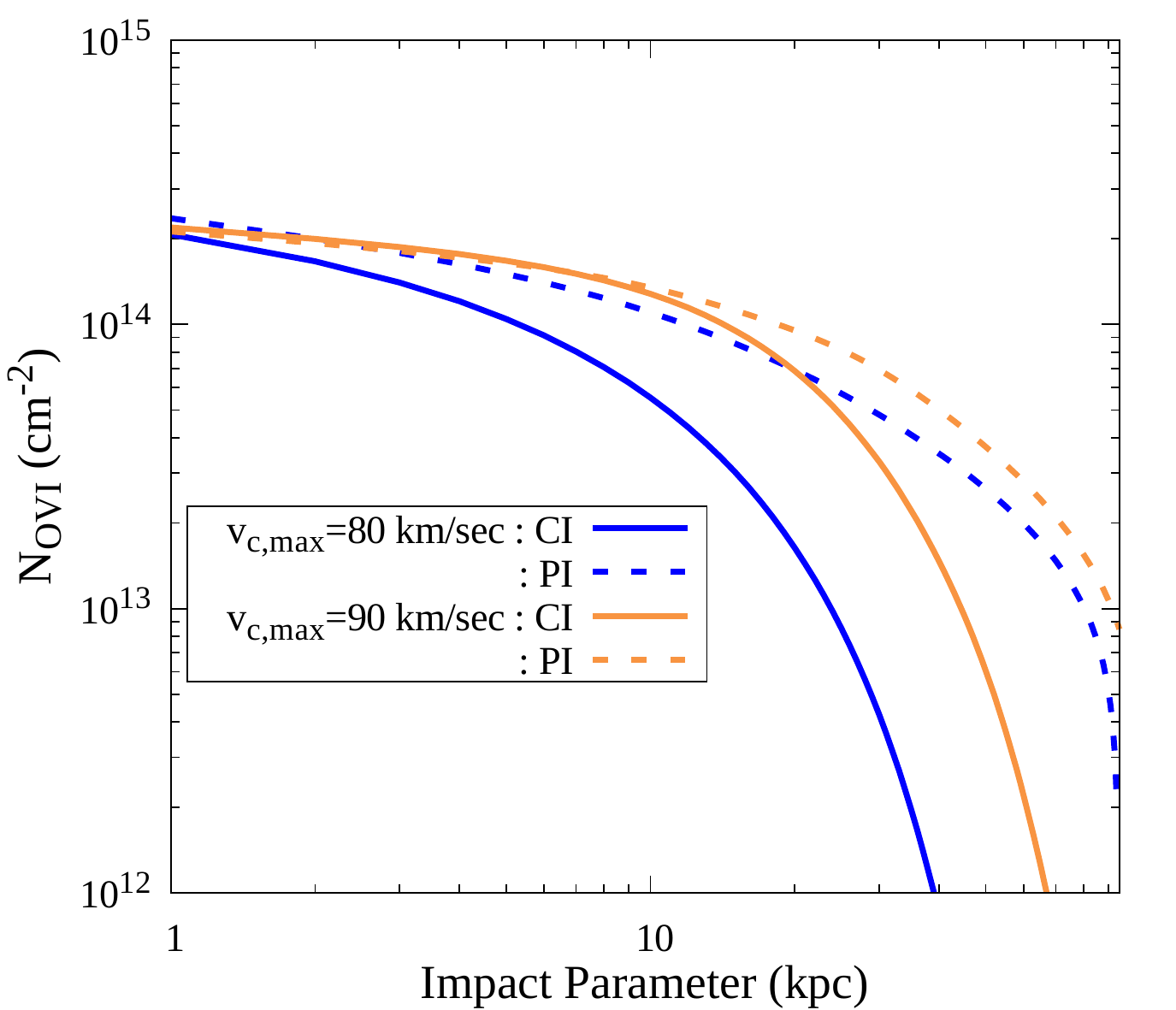}
\includegraphics[width=0.45\linewidth]{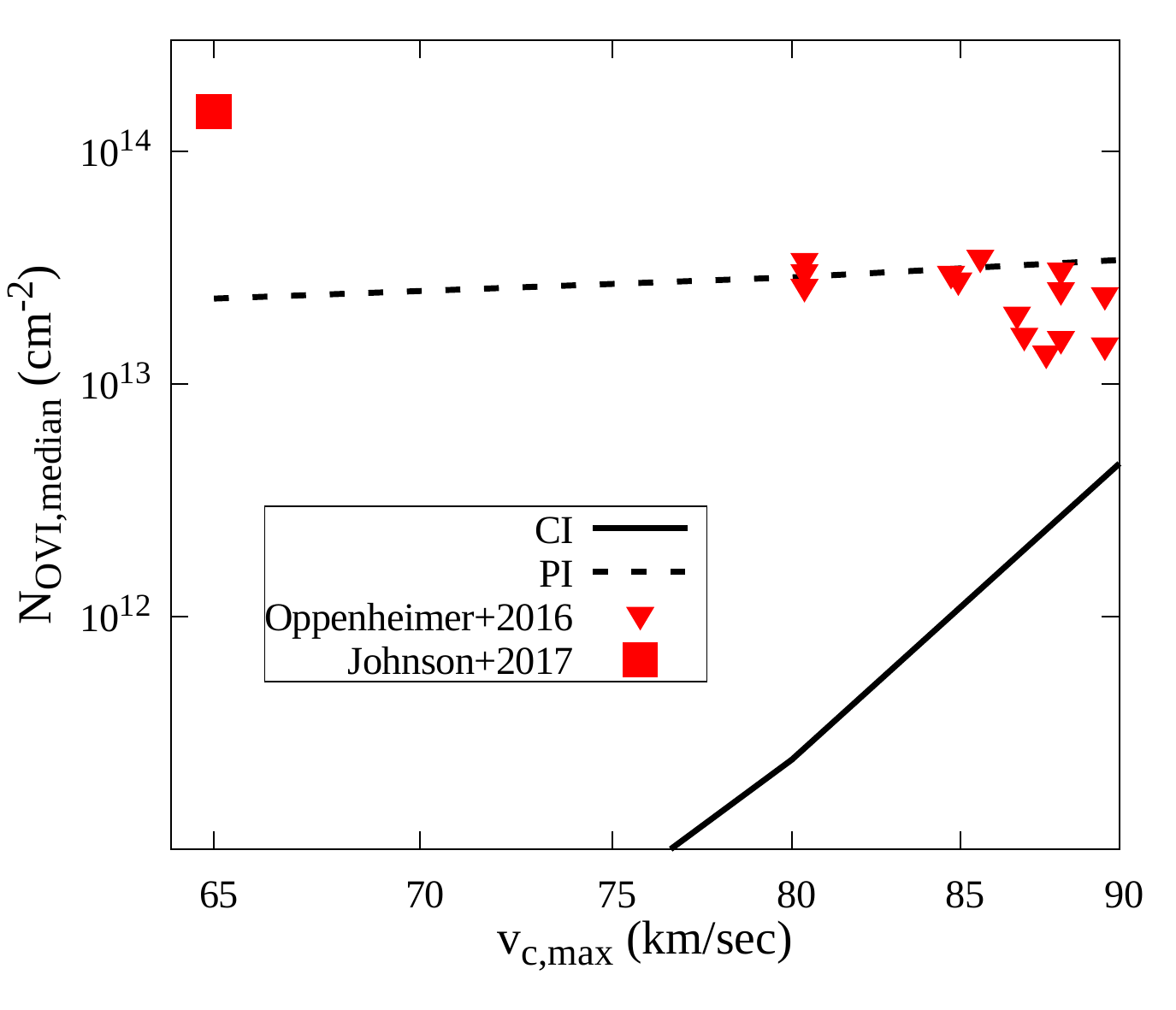}
    \caption{Left panel shows the variation of OVI column density in case of CI and PI  in the low mass galaxies. The plot in the right panel compares variation of N${_\text{OVI,median}}$ in both CI and PI with the simulated and observed values of N${_\text{OVI}}$ of \citet{Oppen2016} and \citet{John2017}.}
    \label{OVI_lowmass}
  \end{figure*}

{{The left panel of }figure \ref{OVI_lowmass} shows the variation of column densities of OVI with impact parameter in precipitation-limited models of low-mass halos with and without photoionization.}  The curves in this figure show that $N_{\rm OVI}$ {greatly} increases from its CI value when photoionization is included. The plot {in the right panel} shows that with the inclusion of photoionization, N$_{\rm{OVI}}$ tends to agree with the simulated values in \cite{Oppen2016} of OVI column density, while the CI value does not. However, the PI value (integrated out to R$_\text{{vir}}$) is not quite as large as the value of N$_{\rm{OVI}}$ observed by \cite{John2017} in dwarf galaxies. However, if we extend the OVI column-density integration to $1.2$R$_\text{{vir}}$ due to the uncertainty in the CGM extent, the result then agrees with the observations of \cite{John2017}.

\begin{figure*}
    \includegraphics[width=0.45\textwidth]{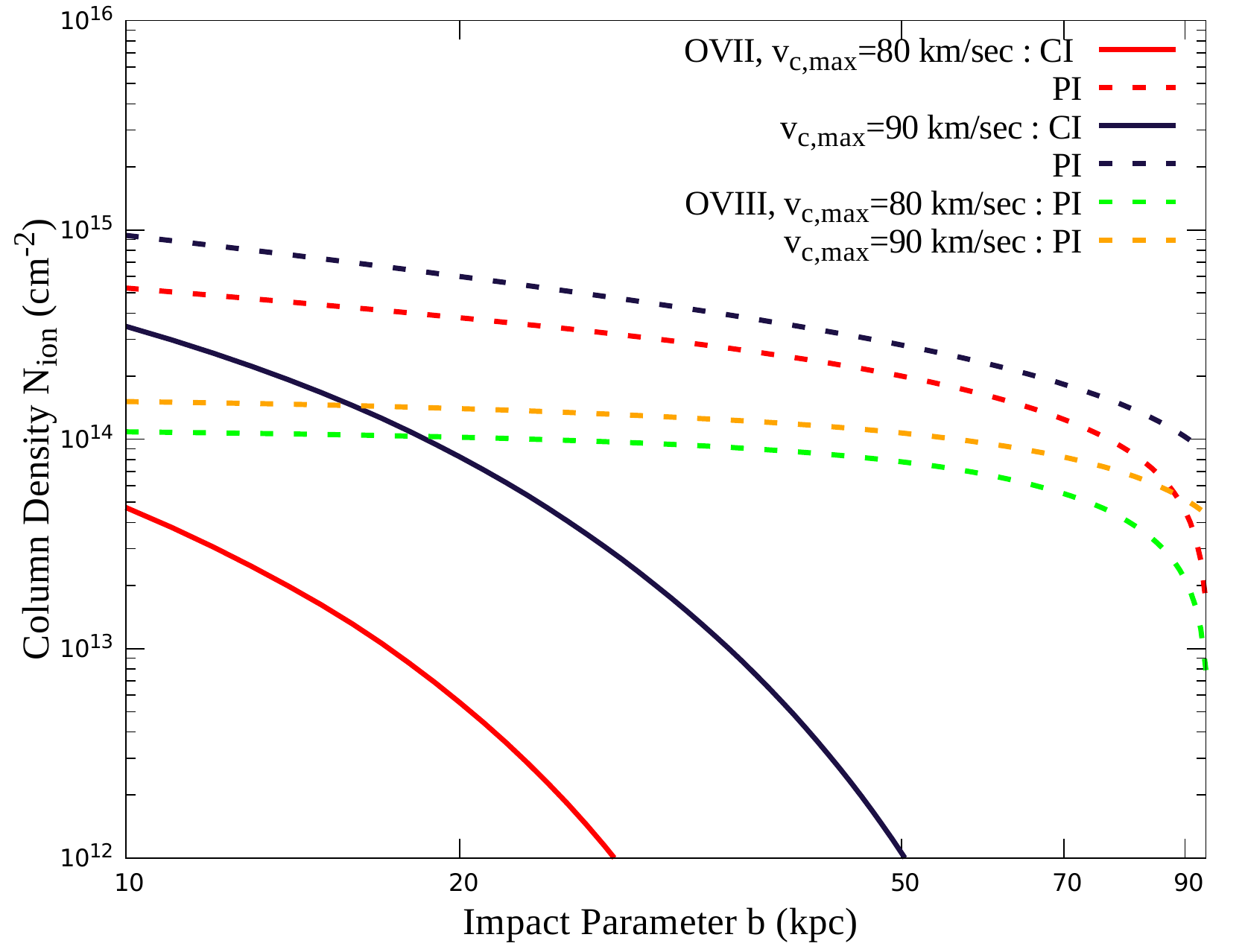}
    \includegraphics[width=0.45\textwidth]{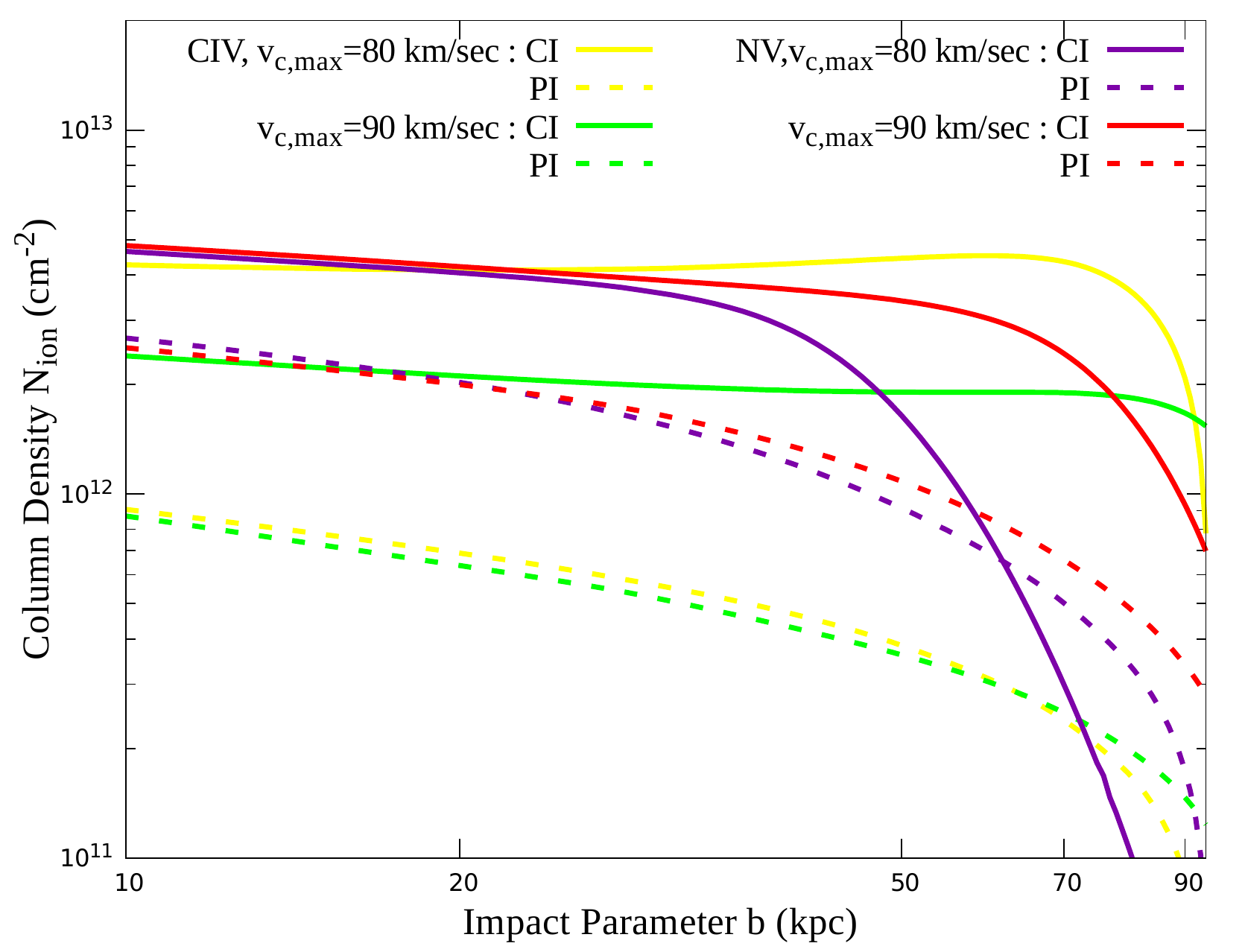}
    \caption{Left and right panels show the variation of OVII, OVIII and NV,CIV column densities respectively in the case of CI and PI in the low mass galaxies.}
    \label{lowmass}
  \end{figure*}

{{The left panel of} figure \ref{lowmass} shows that} $N_{\rm OVII}$ also increases with the inclusion of photoionization. However, the largest effect of photoionization is on $N_{\rm OVIII}$, which does not produce any measureable absorption in the CI case because the CI temperature peak for OVIII ion is considerably different from the virial temperature of a low-mass galaxy. Photoionization is required to produce significant amounts of OVIII. Thus, for low mass galaxies, $N_{\rm OVIII}$ can be a possible diagnostic for the effects of photoionization.

At the same time we find that CIV density decreases when photoionization is included {(in the right panel of Figure \ref{lowmass}),} and it is not possible for this model to match COS-Halos observations of large CIV column densities in dwarf galaxies \citep{Bor2014}, who found $N_{\rm CIV}\sim 10^{14}$ cm$^{-2}$ in such galaxies. {However, there are only upper limits on NV for external galaxies.} There can be several reasons for the mismatch between observed and predicted CIV column densities. Gas in low mass galaxies may deviate significantly from hydrostatic equilibrium \citep{Lo2020}. Also, the gas particle density in galaxies with
$v_\text{c}\le 100$ km s$^{-1}$ dips down to $\sim 10^{-6}$ cm$^{-3}$ beyond $50$ kpc. Given the recombination coefficients of CV (a few times $10^{-12}$ cm$^3$ s$^{-1}$), the recombination time scale becomes Gyr or more. This contrasts with the larger galaxies considered in previous sections, where densities in the outskirts are of order
$10^{-4}$ cm$^{-3}$, and the recombination time scale is of order $\sim 100$ Myr. This implies non-equilibrium
cooling in low mass galaxies, which means that equilibrium calculations may not succeed in explaining observations. {For low mass galaxies, the cooling time in the central region is much shorter than 1 Gyr ($\sim0.4$ Gyr for $v_{c,max}\sim 90$ km s$^{-1}$). Gas in this region is far from hydrostatic equillibrium and there can be a cooling flow in the inner CGM. However, the outer CGM, at say $50$ kpc, where the column density calculations have been done, the cooling time is $\sim 2.2$ Gyr. Therefore our calculations are valid at least for this time scale. 
There is certainly a scope for refinement in this model in future by considering deviation from hydrostatic equilibrium by taking into account cooling flow or wind ejection in the inner CGM.} However, our aim here has been to study the importance of photoionization in low-mass galaxies, which seems to be quite significant, {in anticipation of future observations.}

\section{CONCLUSION}
\label{CONCLUSION}
We extend the precipitation model of \cite{Voit2019} for CGM and investigate the Galactic latitude and longitude variation of column density of OVI, OVII and OVIII. We include photoionization in the model to determine its {effects on both} Milky-way type and low mass galaxies. Our results are summarized below: 

\begin{itemize}

\item { Our precipitation} model can {account for} OVIII observations {of} the Milky Way CGM for $\sigma_{\ln T}\sim 0.6\hbox{--}1.0$, and OVI observations for $\sigma_{\ln T}\sim 1.0$. The {indicated} ratio of OVII to OVIII column density {depends on whether or not we take the median of ratios along individual sight lines (which gives $\sigma_{\ln T} \approx 1.0$) or ratio of the medians (which gives $\sigma_{\ln T} \approx 0.6$).} This range is broadly consistent with the previous findings of \cite{Voit2019} who considered OVI column density for broad absorbers at different impact parameters of other galaxies and estimated $\sigma_{\ln T}\sim 0.7$.

\item  {Photoionization does not play a significant role in CGM of Milky-way type galaxies. This is because of the fact of that the typical CGM temperatures in this case lie in the range favourable for collisional ionization of OVI, OVII, OVIII ions, {if there is} a wide (log-normal) distribution of temperature.}

\item {However, photoionization has a significant effect in the case of low mass galaxies, {in which the virial temperature} is far from the collisional peak for OVI, OVII and OVIII ions. The calculated values from the photoionized precipitation model {are similar to} the observed OVI column density of low mass galaxies. The largest effect of photoionization for low mass galaxies is seen in the column density of OVIII. Hence, OVIII observations can be a probe for the effect of photoionization in low mass galaxies.}

\item {Our precipitation model} implies that star formation related processes in {a halo's central galaxy are} likely correlated with temperature fluctuations in its CGM.  {Comparing the model with observations indicates} temperature fluctuations $\sigma_{\ln T} \approx 1.0$ {in the CGM around} star-forming galaxies. 

\end{itemize}

\bigskip
\section*{Acknowledgement}
We wish to thank Yakov Faermann, Smita Mathur and Anjali Gupta for useful discussions on the subtleties of column density measurements. We also thank the anonymous referee for the detailed comments which have been helpful to improve the manuscript.

\section*{Data Availability}
The data underlying this article are available in the article.

\begin{footnotesize}
\bibliographystyle{mnras}
\bibliography{reference}
\end{footnotesize}


\bsp
\label{lastpage}
\end{document}